\documentclass{llncs}

\usepackage{graphicx}
\usepackage{amsmath}
\usepackage{amssymb}
\usepackage{color}
\usepackage{tabularx}
\usepackage[hidelinks]{hyperref}
\usepackage{url}
\usepackage[T1]{fontenc} 
\usepackage[utf8]{inputenc}

\usepackage{hyphenat}
\usepackage[english]{babel}

\usepackage{mathtools}
\usepackage{subcaption}

\usepackage{algorithm}
\RequirePackage[noend]{algpseudocode}
\usepackage{listings}
\usepackage{wrapfig}
\usepackage[section]{placeins}
\usepackage{stmaryrd}

\DeclareMathOperator{\ite}{\texttt{ite}}

\DeclareMathOperator{\umul}{\texttt{umul}}
\DeclareMathOperator{\shl}{\texttt{shl}}
\DeclareMathOperator{\sext}{\texttt{sext}}
\DeclareMathOperator{\bvmul}{\texttt{bvmul}}
\DeclareMathOperator{\bvudiv}{\texttt{bvudiv}}
\DeclareMathOperator{\bvsdiv}{\texttt{bvsdiv}}
\DeclareMathOperator{\bvurem}{\texttt{bvurem}}
\DeclareMathOperator{\bvsrem}{\texttt{bvsrem}}

\newcommand{\qfbv}[0]{\texttt{QF\_BV}}
\newcommand{\qfufbv}[0]{\texttt{QF\_UFBV}}
\newcommand{\sat}[0]{\texttt{sat}}
\newcommand{\unsat}[0]{\texttt{unsat}}
\newcommand*\wc{{\mkern 2mu\cdot\mkern 2mu}}

\RequirePackage{booktabs}
\RequirePackage{longtable}
\RequirePackage{array}
\RequirePackage{tabu}
\RequirePackage{tabularx}
\RequirePackage{siunitx}
\RequirePackage{multirow}
\RequirePackage{rotating}
\title{An Incremental Abstraction Scheme for Solving Hard SMT-Instances over Bit-Vectors}
\author{Samuel Teuber, Marko Kleine Büning, Carsten Sinz}
\institute{
 Karlsruhe Institute of Technology (KIT), Germany\\
 \url{samuel.teuber@student.kit.edu, {marko.kleinebuening, carsten.sinz}@kit.edu}
}

\begin{document}

\maketitle
\begin{abstract}
Decision procedures for SMT problems based on the theory of bit-vectors are a fundamental component in state-of-the-art software and hardware verifiers. 
While very efficient in general, certain SMT instances are still challenging for state-of-the-art solvers (especially when such instances include computationally costly functions).
In this work, we present an approach for the quantifier-free bit-vector theory (\qfbv{} in SMT-LIB) based on incremental SMT solving and abstraction refinement.
We define four concrete approximation steps for the multiplication, division and remainder operators and combine them into an incremental abstraction scheme.
We implement this scheme in a prototype extending the SMT solver Boolector and measure both the overall performance and the performance of the single approximation steps.
The evaluation shows that our abstraction scheme contributes to solving more unsatisfiable benchmark instances, including seven instances with unknown status in SMT-LIB.
\end{abstract}

\section{Introduction}
\label{ch:Introduction}

Decision procedures for bit-vectors play an important role in many applications such as
bounded model checking, property directed reachability, test generation for hardware circuits,
or symbolic execution \cite{ESBMC12,Gurfinkel-etal:14,LLBMC12,Peleska-etal:11}.
These applications can result in formulae of considerable size, and although
many of them are still within the reach of current implementations of SMT solvers, even
some small formulas remain extremely hard to solve (e.g., the \texttt{modmul} instances from 
the LLBMC Family of Benchmarks \cite{Iser-Kutzner-Sinz-LLBMC-Bench}). It is well known that particular operators
in the logic of bit-vectors possess only large SAT encodings, and thus make problems containing
them often hard to solve. This includes the operators of multiplication, division and remainder.

A technique that is frequently employed to speed up the solving process for hard instances
is \emph{abstraction} \cite{BrummayerBiere:09,p6-brummayer,lahiri2007interpolant,Bryant2007_Chapter_DecidingBit-VectorArithmeticWi-UCLID,MathSAT5,ACDCL,MCSat}. Instead of the original problem, a related problem is analyzed that
is supposed to be easier to solve. Abstractions include under-approximations (where the abstract
system allows for fewer solutions than the original one) and over-approximations (more solutions).

We first present a general scheme for replacing hard bit-vector operators by
a series of over-approximations. Then, to demonstrate and analyze their applicability, we
instantiate the scheme for the operators of multiplication, division and remainder.
The abstraction sequence contains approximations of differing precision, which are tried in turn.
First, less precise approximations are applied and, if they are not sufficient, refined by including 
additional elements of the abstraction sequence. When and which refinements are tried is computed
in a CEGAR-like fashion \cite{CEGAR}.

We have enhanced the SMT solver Boolector \cite{NiemetzPreinerBiere-JSAT15} resulting in an
implementation we call \emph{Ablector}. For an evaluation, we have used the relevant subset of
benchmarks from the SMT-LIB \cite{BarFT-SMTLIB} benchmark release 2019-05-20.
In comparison to Boolector, Ablector is able to solve 11 unsatisfiable instances and 7 instance with unknown status more and yields a total of 46 uniquely solved instances with unsatisfiable or unknown status in SMT-LIB.
Compared to previous work, we consider the use of multiple abstractions and the development of a general,
multi-level abstraction scheme coupled with a strategy for its evaluation and analysis as the main contribution of this paper.

\subsubsection{Related Work.} In the past, a variety of abstraction techniques have been
proposed to speed up SMT solvers.
De Moura and Rueß \cite{DeMouraRuess:2002} presented an approach called ``lemmas on demand'', in which formulas are converted to Boolean constraints, which are then iteratively refined by adding lemmas
generated by the theory solver.
Brummayer and Biere \cite{BrummayerBiere:09,p6-brummayer}
applied this technique to the theory of bit-vectors and arrays, and combined it with under-approximations via bitwidth-reduction. Lahiri and Mehra \cite{lahiri2007interpolant} developed an algorithm combining under- and
over-approximations for the theory of quantifier-free Presburger arithmetic (QFP) based on interpolation.
Bryant \textit{et al.} \cite{Bryant2007_Chapter_DecidingBit-VectorArithmeticWi-UCLID} present an approach
where formulae in bit-vector logic are encoded with fewer Boolean variables than their width, resulting
in an under-approximation. If the under-approximation is unsat\-isfiable, they compute an unsatisfiable core
to derive an over-approximation, which then, in turn, is used to refine another under-approximation.
In the MathSAT solver \cite{MathSAT5} an over-approximating preprocessing step is employed, which 
treats bit-vector operations as uninterpreted functions.
In his PhD thesis proposal, Jonas \cite{Jonas:16} gives a summary of different techniques used in current
bit-vector SMT solvers.
Finally, Brain \textit{et al.} \cite{ACDCL} developed a general framework for abstraction that
generalizes the CDCL algorithm for SAT solving to more expressive theories.
A similar idea was presented around the same time by de Moura and Jovanovic \cite{MCSat}.

\section{Preliminaries}
\label{ch:Preliminaries}
We use common notation for propositional logic and many-sorted first-order logic as can be reviewed in \cite{Barrett-Tinelli-SMT,Marques-Silva-PropositionalSATSolving}.
In particular, we define a signature as $\Sigma = \left(\Sigma^S, \Sigma^P, \Sigma^F, \mathcal{s}^P, \mathcal{s}^F \right)$ where $\Sigma^S$ are the available \textit{sorts}, $\Sigma^P$ are the \textit{predicate symbols}, $\Sigma^F$ are the \textit{function symbols} and  $\mathcal{s}^P\colon \Sigma^P \to \left(\Sigma^S\right)^\ast$ ($\mathcal{s}^F\colon \Sigma^F \to \left(\Sigma^S\right)^+$) defines the \textit{rank} of a given predicate (function).
We call the number of input parameters of a predicate (function) its \textit{arity}.
Furthermore, we denote a $\Sigma$-interpretation as follows:

\begin{definition}[$\Sigma$-interpretation]
    For a signature $\Sigma$ and a set $X$ of variables with sorts in $\Sigma^S$, a \textit{$\Sigma$-interpretation} over $X$ is a tuple
    $\mathcal{I}=\left(\mathcal{U},\mathcal{I}^S, \mathcal{I}^X, \mathcal{I}^F, \mathcal{I}^P\right)$
    where:
    \begin{itemize}
        \item $\mathcal{U}\neq\emptyset$ is the universe of all possible values;
        \item $\mathcal{I}^S\colon\Sigma^S\to\mathcal{P}(\mathcal{U})$\footnote{$\mathcal{P}\left(S\right)$ is the powerset over $S$} maps each sort $\sigma_i$ to a pairwise disjunct domain $D_i\coloneqq\mathcal{I}^S\left(\sigma_i\right)$ of possible values for $\Sigma$-terms of this sort;
        \item $\mathcal{I}^X\colon X\to \mathcal{U}$ maps each variable $x\in X$ to a value $v\in\mathcal{U}$;
        \item $\mathcal{I}^F$ maps any function symbol $f\in\Sigma^F$ of rank  $\mathcal{s}^F\left(f\right)=\sigma_1\dotsi\sigma_n\sigma_{n+1}$ to a function $f^\mathcal{I}\colon D_1\times\dotsi\times D_n \to D_{n+1}$; and
        \item $\mathcal{I}^P$ maps any predicate $p\in\Sigma^P$ of rank $\mathcal{s}^P\left(p\right)=\sigma_1\dotsi\sigma_n$ to a truth function $p^\mathcal{I}\colon D_1\times\dotsi\times D_n \to \{0,1\}$.
    \end{itemize}
    $\mathcal{I}^X$ must respect the sort $\sigma_i$ of $x$ (i.e., $x$ of sort $\sigma_i$ may only be mapped to $v\in D_i$).
\end{definition}

A $\Sigma$-interpretation $\mathcal{I}$ is a \textit{$\Sigma$-model} for some formula $\phi$ iff $\mathcal{I}$ \textit{satisfies} $\phi$ (i.e., $\mathcal{I}\vDash\phi$).
Based on this, a \textit{$\Sigma$-Theory} is a tuple $T=\left(\Sigma,A\right)$ where $\Sigma$ is a signature and $A$ is a set-theoretical class of $\Sigma$-interpretations.
Furthermore, we denote $\textsc{For}^1_{\texttt{T}}$ as all formulae in first-order logic in theory \texttt{T} and $\textsc{Term}^\Sigma_{\sigma_1}$ as the set of all terms in first-order logic with signature $\Sigma$ of sort $\sigma_1$. Given some term $t\in\textsc{Term}_\sigma^\Sigma$, we define
$\phi\left[\texttt{op}\left(\overline{x}\right)\mapsto t \right]$ as the formula where the function application $\texttt{op}\left(\overline{x}\right)$ is replaced by $t$ in $\phi$ (note that $\overline{x}$ represents the vector of all input values for $\texttt{op}$).

In the SMT-LIB standard \cite{BarFT-SMTLIB} for \texttt{QF\_BV}, the functions examined in this work support \textit{overloading} in the sense that a single function symbol like $\bvmul$ supports multiple ranks.
To simplify the explanations in the following sections, one can think of $\bvmul^r$ as the $\bvmul$ operation of rank $r$, thereby avoiding the issue of overloading. This way every function symbol has exactly one rank.
Finally, in \qfbv, we denote $x[i]$ as the $i$th bit of some bit-vector $x$ counting from zero and $x[0]$ is the least significant bit. Further, we denote $x[i:j]$ with $i>j$ as the slice from the $j$th to the $i$th bit of said bit-vector.

\section{Abstraction Scheme}
\label{ch:Scheme}
We present an abstraction procedure for the quantifier-free bit-vector theory (\qfbv).
Our approach substitutes applications of specific operators (for this work specifically $\bvmul$, $\bvsdiv$, $\bvudiv$, $\bvsrem$ and $\bvurem$) by \textit{abstractions} defined on the \qfufbv{} theory (adding uninterpreted
functions to the theory of bit-vector).
During the solving process, the abstractions made within some instance are being iteratively refined until the SAT/SMT solver either returns \unsat{}, or \sat{} with correct assignments.
We will present a formal definition of our  scheme starting with the approximation of some given function symbol:

\begin{definition}[Approximation]
    \label{def:Scheme:Approximation}
    Given some theory $T=\left(\Sigma,A\right)$ and some function symbol $\texttt{op}\in\Sigma^F$ with $\mathcal{s}^F\left(\texttt{op}\right)=\sigma_1\dotsi\sigma_n\sigma$ and $n\geq1$, a $T$-approximation for \texttt{op} consists of:
    \begin{itemize}
        \item a new uninterpreted function symbol $ap_{\texttt{op}}$ with $\mathcal{s}^F\left(\texttt{op}\right) = \mathcal{s}^F\left(ap_{\texttt{op}}\right)$; and
        \item a mapping $\mathcal{A}_{\texttt{op}}\colon \textsc{Term}^\Sigma_{\sigma_1}\times\dotsi\times\textsc{Term}^\Sigma_{\sigma_n} \to \mathcal{P}({\textsc{For}^1_{\texttt{T}}})$.
    \end{itemize}
    A $T$-approximation can therefore be written as a tuple $\left(ap_{\texttt{op}}, \mathcal{A}_{\texttt{op}}\right)$.
\end{definition}

An approximation essentially replaces an occurrence of an existing function symbol $\texttt{op}$ by a new one ($ap_{\texttt{op}}$), and furthermore adds formulae that ensure certain properties for the application of $ap_{\texttt{op}}$.
It may be \textit{sound} or \textit{complete}:

\begin{definition}[Sound $T$-approximation]
    \label{def:Scheme:Sound}
Given some theory $T=\left(\Sigma,A\right)$, a $T$-approximation $\left(ap_{\texttt{op}}, \mathcal{A}_{\texttt{op}}\right)$ is \textit{sound} iff for all
$\overline{x}\in Dom \left(\mathcal{A}_{\texttt{op}}\right)$ the following property holds:\footnote{$Dom$ is the domain of a given function. In this case $Dom\left(\mathcal{A}_{\texttt{op}}\right) = \textsc{Term}^\Sigma_{\sigma_1}\times\dotsi\times\textsc{Term}^\Sigma_{\sigma_n}$.}
For all $T+UF$-interpretations $\mathcal{I}$ with $\mathcal{I}\vDash\mathcal{A}_{\texttt{op}}\left(\overline{x}\right)$, it holds that
$\mathcal{I}\vDash ap_{\texttt{op}}\left(\overline{x}\right) \doteq \texttt{op}\left(\overline{x}\right)$.
\end{definition}

\begin{definition}[Complete $T$-approximation]
\label{def:Scheme:Complete}
Given some theory $T=\left(\Sigma,A\right)$, a $T$-approximation $\left(ap_{\texttt{op}}, \mathcal{A}_{\texttt{op}}\right)$ is \textit{complete} iff for all $\overline{x}\in Dom\left(\mathcal{A}_{\texttt{op}}\right)$ the following property holds:
For all $T+UF$-interpretations $\mathcal{I}$ with $\mathcal{I}\vDash ap_{\texttt{op}}\left(\overline{x}\right) \doteq \texttt{op}\left(\overline{x}\right)$, it holds that
$\mathcal{I}\vDash\mathcal{A}_{\texttt{op}}\left(\overline{x}\right)$.
\end{definition}
A sound $T$-approximation is an under-approximation, while a complete $T$-ap\-prox\-i\-ma\-tion is an over-ap\-prox\-i\-ma\-tion of some function $\texttt{op}$.
A set of approximations can then be used to construct an \textit{abstraction scheme}:
\begin{definition}[Abstraction Scheme]
    \label{def:Scheme:Abstraction}
    Given some theory $T=\left(\Sigma,A\right)$ and some function symbol $\texttt{op}\in\Sigma^F$ of strictly positive arity, a $T$-abstraction scheme (for $\texttt{op}$) is a finite, totally ordered set of $\ T$-approximations \[
    \mathcal{AS}_{\texttt{op}} = \{ \left(ap_{\texttt{op}}, \mathcal{A}^1_{\texttt{op}}\right),\dots,\left(ap_{\texttt{op}}, \mathcal{A}^k_{\texttt{op}}\right) \}
    \vspace{-1ex}
    \]
    where:
    \begin{itemize}
        \item For every $i\in\llbracket1,k\rrbracket$: $\left(ap_{\texttt{op}},\mathcal{A}^i_{\texttt{op}}\right)$ is a complete $T$-approximation of \texttt{op} and
        \item $\left(ap_{\texttt{op}},\mathcal{C}_{\texttt{op}}\right)$ with $\mathcal{C}_{\texttt{op}}\left(\overline{x}\right)\coloneqq\!\!\!
        \bigcup\limits_{\left(\wc,\mathcal{A}\right)\in\mathcal{AS}_{\texttt{op}}} \!\!\! \mathcal{A}\left(\overline{x}\right)$ is a sound $T$-approximation of \texttt{op}.\footnote{
            Just like all previous $T$-approximations, $\mathcal{C}_{\texttt{op}}$ is defined as
            $\mathcal{C}_{\texttt{op}}\colon \textsc{Term}^\Sigma_{\sigma_1}\times\dotsi\times\textsc{Term}^\Sigma_{\sigma_n} \to \mathcal{P}({\textsc{For}^1_{\texttt{T}}})$ for a function symbol \texttt{op} of rank $\sigma_1\dotsi\sigma_n\sigma$.
        }
    \end{itemize}
\end{definition}

While any single approximation within the abstraction scheme is only complete (and therefore an over-approximation), all approximations taken together must be sound and should thus yield a correct definition of the original function\footnote{Theorems and Proofs for the correctness of the abstraction scheme can be found in Appendix \ref{sec:appendix:proof}}.

This abstraction scheme can then be used to build a decision procedure like the one described in Algorithm \ref{algorithm:Scheme:Refinement}.
In a first step, the algorithm replaces all operators which should be refined by their abstracted uninterpreted functions. Afterwards, the instance is re-evaluated in a loop as long as the underlying SMT solver does not return \texttt{unsat} and the model returned is incorrect. In each round, further approximations from the abstraction schemes are added to the instance.
This process is certain to converge once all approximations of the scheme have been added and the formulation is thus both sound and complete.

\begin{algorithm}[]
    \caption{Decision procedure for QF\_BV abstractions. \texttt{ADD\_CLAUSES} and \texttt{SAT} are calls to the underlying SMT solver.}
    \label{algorithm:Scheme:Refinement}
    \begin{algorithmic}
    \Require $\phi \in \textsc{For}^1_{\qfbv}$
    \State $functions \gets \langle\rangle$
    \For{$\texttt{op}\in\Sigma^F$}
         \If{\texttt{op} has abstraction}
         \Comment{For $\bvmul$, division and remainder in our case}
             \For{$\texttt{op}\left(\overline{x}\right)$ in $\phi$}
             \Comment{For each occurrence of \texttt{op}}
                 \State $\phi \gets \phi\left[\texttt{op}\left(\overline{x}\right) \mapsto ap_\texttt{op}\left(\overline{x}\right)\right]$
                 \State $functions.push\left( \left(\mathcal{AS}_\texttt{op}, \texttt{op}\left(\overline{x}\right) \right)\right)$
             \EndFor
         \EndIf
    \EndFor
    \State \texttt{ADD\_CLAUSES($\phi$)}
    \While{true}
        \State $r \leftarrow $\texttt{SAT()}
        \Comment{Checks for satisfiability of current instance}
        \If{$r=$\unsat}
             \State \textbf{return} \unsat
             \Comment{Correct result was found}
        \Else
             \State $consistent \gets true$
             \For{$\left(\mathcal{AS}, \texttt{op}\left(\overline{x}\right) \right)$ in $functions$}
                 \If{$\texttt{op}\left(\overline{x}\right)$ assignment is inconsistent}
                 \Comment{Abstraction needs refinement}
                    \State $consistent \gets false$
                    \State \texttt{ADD\_CLAUSES($\mathcal{AS}.pop()$)}
                    \Comment{Adds next abstraction step to instance}
                 \EndIf
             \EndFor
             \If{consistent}
                \State \textbf{return} \sat
                \Comment{Correct result was found}
             \EndIf
        \EndIf
    \EndWhile
    \end{algorithmic}
\end{algorithm}

\section{Abstractions}
\label{ch:Abstractions}
We developed abstraction schemes for the computationally costly functions $\bvmul$, $\bvsdiv$, $\bvudiv$, $\bvsrem$ and $\bvurem$. We present the abstraction of $\bvmul$ in more detail, while giving just a short overview of the $\bvsrem$ abstraction. We are omitting a description of the other functions due to similarity and space limitations. 
\subsection{Abstracting $\bvmul$}
\label{subsec:Abstractions:bvmul}
The abstraction scheme for $\bvmul$ is divided into four stages:
The first stage describes the behavior of $\bvmul$ for various simple cases (like factors 0 and 1);
the second stage defines intervals for the result value given the intervals of the multiplication factors;
the third stage introduces relations between $\bvmul$ and other functions (specifically division and remainer) and
the fourth stage finally adds full multiplication for certain intervals of the factors.

Throughout the abstraction process, we will consider $\bvmul$ a signed operation, that is, we will interpret $r=\bvmul\left(x,y\right)$ as if $x$, $y$ and $r$ were signed integers.
Even though this seems like a restriction, we do not lose correctness of our approach for unsigned values doing this.
For example, if we assert an over-approximation (for an 8 bit mul\-ti\-pli\-ca\-tion) like
$x <_s 0_s \land y <_s 0_s \implies r >_s 0_s$,
this over-approximation also holds for unsigned values as it can be interpreted as
$x \ge_u 128_u \land y \ge_u 128_u \implies r <_u 128_u$ for unsigned values.
While this might be a surprising abstraction, it is nonetheless a correct one.\footnote{For simplicity, we are omitting some overflow behavior in this example. However, we do consider overflow cases in the approximations defined later on.}
Effectively, the question of whether $x$, $y$ and $r$ are signed or unsigned, is an issue for the user's interpretation and not for the decision procedure itself.

\paragraph{Overflow detection.}
For many of the abstractions proposed in this chapter, it is essential to detect overflows of $\bvmul$.
To this end, we defined a predicate $noov: \{0,1\}^w \times \{0,1\}^w \rightarrow \{0,1\}$ for bitwidth $w$ based on \cite{Warren-HackersDelight} which is true if an overflow can happen when multiplying the two input variables.
The approach works by counting the number of leading bits (ones or zeros).
Note that this predicate also detects signed overflows and
might not be sound\footnote{While every overflow will be detected,
it might detect more overflows than actually exist}.


\subsubsection{Simple cases.}
\label{subsubsec:Abstractions:bvmul:simple}
For a multiplication instance $\bvmul\left(x,y\right)$ of factors $x$ and $y$ with bitwidth $w$, we define the following constraints:
\begin{flalign}
    \qquad\qquad\qquad
    \left(x \doteq 0\right) &\Rightarrow \qquad \left(ap_{\bvmul}\left(x,y\right) \doteq 0\right)
        &\label{align:Abstractions:bvmul:simple:zero1}\\
    \left(y \doteq 0\right) &\Rightarrow \qquad \left(ap_{\bvmul}\left(x,y\right) \doteq 0\right)
        &\label{align:Abstractions:bvmul:simple:zero2}\\
    \left(x \doteq 1\right) &\Rightarrow \qquad \left(ap_{\bvmul}\left(x,y\right) \doteq y\right)
        &\label{align:Abstractions:bvmul:simple:one1}\\
    \left(y \doteq 1\right) &\Rightarrow \qquad \left(ap_{\bvmul}\left(x,y\right) \doteq x\right)
        &\label{align:Abstractions:bvmul:simple:one2}\\
    \left(x \doteq -1\right) &\Rightarrow \qquad \left(ap_{\bvmul}\left(x,y\right) \doteq -y\right)
        &\label{align:Abstractions:bvmul:simple:neg1}\\
    \left(y \doteq -1\right) &\Rightarrow \qquad \left(ap_{\bvmul}\left(x,y\right) \doteq -x\right)
        &\label{align:Abstractions:bvmul:simple:neg2}
\end{flalign}
\begin{flalign}
    noov(x,y) \Rightarrow \qquad&
        \left( \neg x[w-1] \land \neg y[w-1] \right)
            &\Rightarrow&\quad
            \neg ap_{\bvmul}\left(x,y\right)[w-1]
                &\label{align:Abstractions:bvmul:simple:bothPos}\\
            \land & \left( x >_s 0 \land y[w-1] \right)
            &\Rightarrow&\quad
            ap_{\bvmul}\left(x,y\right)[w-1]
                &\label{align:Abstractions:bvmul:simple:oneNeg1}\\
            \land & \left( x[w-1] \land y  >_s 0 \right)
            &\Rightarrow&\quad
            ap_{\bvmul}\left(x,y\right)[w-1]
                &\label{align:Abstractions:bvmul:simple:oneNeg2}\\
            \land  & \left( x[w-1] \land y[w-1] \right)
            &\Rightarrow&\quad
            \neg ap_{\bvmul}\left(x,y\right)[w-1]
                &\label{align:Abstractions:bvmul:simple:bothNeg}
\end{flalign}
For example, equations (\ref{align:Abstractions:bvmul:simple:zero1}) and (\ref{align:Abstractions:bvmul:simple:zero2}) define the multiplication cases where one factor is zero, other equations cover similar easy cases. Rules (1)--(4) have also been proposed in \cite{Bryant2007_Chapter_DecidingBit-VectorArithmeticWi-UCLID}, the other rules are, to the best of our
knowledge, novel.

Additionally, we can make statements about the result's sign whenever we can be certain that no overflow is going to happen.
For the cases where no overflow happens, the sign behavior of bit-vector multiplication corresponds to the common sign behavior of multiplication and can be encoded as an approximation as seen in (\ref{align:Abstractions:bvmul:simple:bothPos})--(\ref{align:Abstractions:bvmul:simple:bothNeg}).
Finally, all cases where one of the two factors is a power of 2 can be covered by constraints like (\ref{align:Abstractions:bvmul:simple:pow2}) for all $i\in\llbracket 1,w-1 \rrbracket$ and for $x$ and $y$ symmetrically:
\begin{flalign}
    \bigwedge\limits_{j \neq i} \neg x\left[j\right] \land x\left[i\right] \implies \left( \umul\left(x^+_2,y^+_2\right) \doteq \shl\left( y^+, i \right)  \right)
    \label{align:Abstractions:bvmul:simple:pow2}
\end{flalign}
where $\umul$ is the unsigned multiplication function and $x^+_2$ as well as $y^+_2$ are positive, double bitwidth versions of $x$ and $y$ as detailed in the following section. Instruction $\shl$ is the left shift function.

The mapping $\mathcal{A}_{\texttt{op}}$ of this abstraction stage is then the conjunction of 
(\ref{align:Abstractions:bvmul:simple:zero1})--(\ref{align:Abstractions:bvmul:simple:pow2}).
These formulae are no static rewrite rules, but constraints provided to the underlying solver.

\paragraph{Completeness.} The completeness is a direct consequence of the definition given in \ref{def:Scheme:Complete}
as it can be checked that all formulae presented above (which specifically omitted any statements about overflow cases) are implications of $ap_{\bvmul}\left(x,y\right) \doteq \bvmul\left(x,y\right)$.

\subsubsection{Highest bit set based intervals.}
\label{subsubsec:Abstractions:bvmul:msd}
Using the factors' \emph{highest bit set}\footnote{The \emph{highest bit set} of $x$ is $i$ iff $x$ is of the form $0^{n-i-1}\cdot1\cdot\{0,1\}^{i}$}, intervals of the factors can be defined, which in turn can be used to assert intervals of the multiplication's result.
In a first step, the signed multiplication $r\coloneqq ap_{\bvmul}\left(x,y\right)$ is transformed into its unsigned version with doubled bitwidth:
\begin{align*}
x_2^+ &\doteq \ite(x[w-1],\ -\sext\left(x,w\right),\ \sext\left(x,w\right))\,, \\
y_2^+ &\doteq \ite(y[w-1],\ -\sext\left(y,w\right),\ \sext\left(y,w\right))\,,\\
r'_2 &\doteq \ite(x[w-1] \oplus y[w-1],\ -\umul(x_2^+,y_2^+),\ \umul(x_2^+,y_2^+) )\,.
\end{align*}
Instruction $\sext$ is the sign extension function.
By asserting equality of the multiplication result $r$ and $r'_2\left[w-1:0\right]$, it is then possible to reason about the results of $r^+_2\coloneqq \umul(x_2^+,y_2^+)$ through bit shifting:
If $i$ is the highest bit set of $x^+_2$ then $2^i\leq x^+_2 < 2^{i+1}$ and therefore, $2^i*y^+_2 \leq r^+_2 < 2^{i+1}*y^+_2$.
We thus define a predicate $hbs(x,i)$ which is true iff the highest bit set of $x$ is $i$.

The previously presented intuition gives rise to the following abstraction which must distinguish overflow from no-overflow cases.
For this, we will initially use a double bitwidth (i.e., $2*w$ width) unsigned multiplication function.
We then define double bit width lower($L$) and upper($U$) bounds for the result of the multiplication based on the highest bit set as explained above:
\begin{flalign*}
    L(a, b, n)\coloneqq&
    \begin{cases}
        \ite\left( hbs(a,0), b, 0 \right) &, n=0\\
        \ite\left( hbs(a,n), \shl\left(b, n\right), L(a, b, n-1) \right) &, else
    \end{cases}
\\
    U(a, b, n)\coloneqq&
    \begin{cases}
        \shl\left(b,1\right) &, n=0\\
        \ite\left( hbs(a,n), \shl\left(b, n+1\right), U(a, b, n-1) \right) &, else\\
    \end{cases}
\end{flalign*}
We can compare the necessary number of bits depending on the result of $noov$: If an overflow is possible, we must compare the version with $2*w$ bits,
otherwise the $w$ bit version can be used for comparison.

Note that while $L$ and $U$ seem to be recursive functions, they can be unrolled into consecutive $\ite$ statements when adding the bounds to the instance at hand.

The mapping $\mathcal{A}_{\texttt{op}}$ of this approximation step is then the assertion that $r^+_2$ must lie within the bounds given by $L(x^+_2,y^+_2,w-1)$ and $U(x^+_2,y^+_2,w-1)$ (for the necessary bitwidth as explained above).

\paragraph{Completeness.}
Through the distinction between overflow and no-overflow cases the various equations can be regarded as normal multiplication disregarding overflows.
Therefore, it can be checked that these constraints are direct implications of $ap_{\bvmul}\left(x,y\right) \doteq \bvmul\left(x,y\right)$.
This approximation is consequently complete according to definition \ref{def:Scheme:Complete}.

\subsubsection{Relations to other functions.}
\label{subsubsec:Abstractions:bvmul:relations}
Aside from previous abstraction approaches, we can also look at relations between functions -- possibly providing the solver with more high-level information.
This can be useful in cases where relations between multiple function applications already lead to a contradiction.

For the multiplication instruction $ap_{\bvmul}(x_2,y_2)$, with $x_2$ and $y_2$ the double bitwidth ($2*w$) versions of $x$ and $y$, we propose the following abstractions:
\begin{flalign*}
    ap_{\bvmul}(x_2,y_2) \doteq&\ ap_{\bvmul}(y_2,x_2)\,,\\
    x_2 \doteq 0 \lor y_2 \doteq&\ ap_\texttt{bvsdiv}(ap_{\bvmul}(x_2,y_2),x_2)\,,\\
    y_2 \doteq 0 \lor x_2 \doteq&\ ap_\texttt{bvsdiv}(ap_{\bvmul}(x_2,y_2),y_2)\,.
\end{flalign*}
For every bit width $w'<2*w$ which appears in a given problem instance and its abstractions, we can further assert that
\[
    ap_{\bvmul}(x_2,y_2)[w'-1:0] \doteq ap_{\bvmul}(x_2[w'-1:0],y_2[w'-1:0])\,,
\]
\[
    ap_{\bvmul}(x_2,y_2)[w'-1:0] \doteq ap_{\bvmul}(y_2[w'-1:0],x_2[w'-1:0])
\]
and for
\[
    x' \coloneqq \sext\left(x_2\left[\left\lfloor \frac{w'}{2} \right\rfloor:0\right], w'-\left\lfloor \frac{w'}{2} \right\rfloor\right);\ 
    y' \coloneqq \sext\left(y_2\left[\left\lfloor \frac{w'}{2} \right\rfloor:0\right], w'-\left\lfloor \frac{w'}{2} \right\rfloor\right)
\]
we assert that
\[
   ap_{\bvmul}(x',y') \doteq ap_{\bvmul}(y',x')\,,
\]
\[
    y'\doteq 0 \lor x' \doteq ap_\texttt{bvsdiv}\left(ap_{\bvmul}(x',y'),y'\right)\,,
\]
\[
    x'\doteq 0 \lor y' \doteq ap_\texttt{bvsdiv}\left(ap_{\bvmul}(x',y'),x'\right)\,.
\]
Essentially all these relations between various multiplication and division applications are all based
on the semantic of multiplication, division and remainder as used in SMT-LIB and C++ \cite{ISO14882:2011}.\\
The only challenge of this abstraction is to formulate the constraints so that they are complete for overflow cases. For this, we use an approach with doubled bitwidth $2*w$ while encoding the constraints and define $x'$ and $y'$ for every $w'$ in a way that prevents overflows during multiplication.

\paragraph{Completeness.}
The completeness of this abstraction is a direct consequence of all assertions being well-known properties for machine multiplication and division. As we avoid all overflow cases through the use of doubled bitwidth, the properties hold for any input combination.

\subsubsection{Full multiplication.}
\label{subsubsec:Abstractions:bvmul:fullmul}
In a last step, full multiplication on a per-interval basis is added as a constraint.
We assume an SMT instance containing some multiplication $\bvmul\left(x,y\right)$. 
If the instance is still satisfiable after the previous steps, a counterexample is returned.
We then look up the highest bit set $i$ of $x$'s assignment and assert that the multiplication is precise if bit $i$ of $x$ is set to $1$:
\[
    hbs\left(x,i\right) \implies ap_{\bvmul}\left(x,y\right) \doteq \bvmul\left(x,y\right)\,.
\]

\paragraph{Completeness and Soundness.}
For a multiplication of bitwidth $w$ the approximation is complete and it even becomes sound once this assertion has been made
for all $i\in\llbracket 0,w-1\rrbracket$. Note, that the maximum number of necessary refinement steps is bounded by the function's bitwidth $w$.


\subsection{Abstracting \texttt{bvsrem}}
\label{subsec:Abstractions:bvsrem}
Due to its rareness in benchmarks\footnote{Its rareness makes it harder to evaluate the performance of abstractions and abstractions are less likely to have a big impact on the overall performance.} only a single abstraction layer has been added for this operator.
Once again with double bitwidth as explained in Section \ref{subsubsec:Abstractions:bvmul:relations}, we assert the relations between \texttt{bvsrem} and other functions:
\begin{flalign*}
    ap_{\texttt{bvsrem}}\left(x,y\right) \doteq& ap_{\texttt{bvsrem}}\left(x_2,y_2\right)\left[w-1:0\right]\\
    x_2 \doteq& ap_{\texttt{bvmul}}\left( ap_{\texttt{bvsdiv}}\left(x_2,y_2\right), y_2 \right) + ap_{\texttt{bvsrem}}\left(x_2,y_2\right)\,.
\end{flalign*}
In the following refinement step, we add the full remainder constraint:
\[
    ap_{\texttt{bvsrem}}\left(x,y\right) \doteq \texttt{bvsrem}\left(x,y\right)\,.
\]

\section{Evaluation}
\label{ch:Evaluation}
In order to evaluate the performance of the abstraction scheme presented above, we implemented a prototype solver by enhancing Boolector \cite{NiemetzPreinerBiere-JSAT15} (version 3.2).
To increase readability, we will refer to the prototype as \textit{Ablector} which stands for \textit{Ab}stracted Boo\textit{lector}.

\paragraph{Experimental setup.}
\begin{wraptable}[11]{R}{0.5\textwidth}
    \centering
    \raisebox{0pt}[\dimexpr\height-1.2\baselineskip\relax]{
    \begin{tabular}{r|c|c}
        & Contribution & Cost \\ \hline\hline
        Step 1 & \textbf{28} & 24  \\ \hline
        Step 2 & 2 & \textbf{3} \\ \hline
        Step 3 & \textbf{15} & 0 \\ \hline
        Step 4 & 1 & 1 \\ \hline \hline
        Sum & 46 & 28
    \end{tabular}
    }
    \caption{Contributions and Costs of the implemented approximation steps.}
    \label{tab:evaluation:contributions:fullscheme}
\end{wraptable}

As the focus of this work lay on researching which abstractions are effective in solving more \textit{hard} instances and not so much on building an improved solver, the abstraction refinement procedure is built as a layer on top of Boolector.
We will compare the default configuration of Boolector with our prototype based upon on said default configuration in order to quantify the contribution of our abstraction scheme.
To enable a fast evaluation of the abstraction approaches, Ablector is implemented in Python.
The abstraction engine was placed in-between the parser PySMT \cite{pysmt2015} and Boolector's Python interface as Boolector's own Python API is not extensible when parsing SMT files. 
As the parsing system of PySMT is not competitive to the one of Boolector, we solely compare the CPU clock time\footnote{We chose the CPU clock time as it can be measured through the same unified interface in both Python and C with comparable results. As the tasks are heavily CPU-bound with very little IO operations, this measurement can still be considered realistic.} of the solving process.
We do this by comparing Boolector's SMT-LIB \texttt{check-sat} call against the summed up CPU clock time of all invocations to rewritten procedures in Ablector including its own \texttt{check-sat} call.
In particular, our time measurement for Ablector also contains all abstraction refinement procedures.
This will produce a more realistic comparison of the abstraction's performance.
Note that we are even over-approx\hyp{}imating the time Ablector takes in this comparison as we are
adding the operator construction time during parsing, which is not considered for Boolector.
These procedure calls, however, are in most cases negligible in comparison to the time for the \texttt{check-sat} call.
In accordance with the rules of the SMT competition 2018 \cite{SMTCOMP18} the timeout was set to $1200s$ in all experiments\footnote{Further details for reproducibility and on the experimental setup can be found in Appendix \ref{sec:appendix:reproducibility}.}.

\paragraph{Benchmark Selection}
Our work was sparked by an investigation on hard benchmarks in the LLBMC family of benchmarks presented at the SAT Competition 2017 \cite{Iser-Kutzner-Sinz-LLBMC-Bench}.
To ensure that the abstractions do not \textit{overfit}, we decided to evaluate on the larger set of $\qfbv$ benchmarks in the 2019-05-20 SMT-LIB Benchmark release \cite{BarFT-SMTLIB} which were used for the 2019 SMT-Competition \cite{SMTCOMP19}.
This benchmark set contains 14382 instances of satisfiable, 27144 instances of unsatisfiable and 170 instances of unknown status.
We decided to remove all benchmarks not using the abstracted operators $\bvmul, \bvsdiv, \bvudiv, \bvsrem$ or $\bvurem$ as we were mainly interested in the effect of the abstractions implemented.
This resulted in a subset containing 6024 instances of satisfiable, 15849 instances of unsatisfiable and 92 instances of unknown status.
Of those instances, Boolector was unable to solve 758 satisfiable, 152 unsatisfiable and 52 unknown instances.
Boolector solves 9769 of the unsatisfiable instances through its rewriting engine thus leaving 6080 instances to be solved by transformation to SAT clauses.

\paragraph{}
Preliminary experiments showed that our abstraction scheme is mainly helpful when solving unsatisfiable instances while worsening the results for satisfiable instances.
This is congruent with the expectation that over-approximation techniques usually help in speeding up the solver’s runtime
for unsatisfiable SMT-instances, while under-approximation techniques help reducing the runtime for satisfiable instances \cite{Brummayer-PhD}.
For this reason, we will begin our analysis by looking at how our abstraction scheme helps in solving unsatisfiable and unknown\footnote{All instances with unknown status in the benchmark set whose status we managed to find out are unsatisfiable, too.} instances.

\subsection{Contributions of approximation steps}
Given the four, incremental approximation steps presented, it is a natural question to ask which approximation step within the abstraction scheme helps how much in solving the instance at hand.
After some preliminary experiments on the ordering of the approximation steps, we came up with the following sequence of steps:
\begin{enumerate}
    \item Simple cases (see \ref{subsubsec:Abstractions:bvmul:simple}, "Simple cases")
    \item Interval based abstraction (see \ref{subsubsec:Abstractions:bvmul:msd}, "Highest bit set based intervals")
    \item Function relations (see \ref{subsubsec:Abstractions:bvmul:relations}, "Relations to other functions")
    \item Iterative, interval based multiplication (see \ref{subsubsec:Abstractions:bvmul:fullmul}, "Full multiplication") \,.
\end{enumerate}

With the abstraction scheme setup in this sequence, we began running experiments on the unsatisfiable and unknown instances.
In five subsequent experiments we ran Boolector, Ablector with only the first step, Ablector with first and second step, Ablector with first through third step and full Ablector on all benchmarks.
Interestingly, there is no solid progression clearly decreasing the number of unsolved instances in every step.
Instead every step makes a number of instances solvable and another number of instances unsolvable.
We therefore defined both the \textit{contribution} and the \textit{cost} of an approximation step:
\begin{itemize}
	\item  We call the \textit{contribution} of an approximation at position $N$ in an abstraction scheme the number of benchmark instances that are not solved by approximation steps 1 to $N-1$ but are solved using the approximations 1 to $N$. 
	Thus, the {contribution} identifies exactly those benchmark instances which are solved through the $N$th approximation step within the scheme.\\
	\item  In contrast, we define the \textit{cost} of an approximation at position $N$ in an abstraction scheme as the number of benchmark instances that are solved by approximation steps 1 to $N-1$ but are not solved using the approximations 1 to $N$. Thus identifying exactly those benchmark instances which are not solved, because the $N$th approximation step within the scheme was put in place.
\end{itemize}
\begin{wraptable}[11]{R}{0.5\textwidth}
    \raisebox{0pt}[\dimexpr\height-0\baselineskip\relax]{
    \begin{tabular}{cc|c|c|c}
        &&\multicolumn{2}{c|}{Boolector}&\\
        &&unsolved&solved&\\ \hline
        \multirow{2}{*}{Ablector}&unsolved& 113 & 28 & 141 \\ \cline{2-5}
        & solved & 39 & 5904 & 5943 \\ \hline
        & & 152 & 5932 & 6084 \\
    \end{tabular}
    }
    \caption{Number of unsatisfiable instances solved by Boolector and Ablector (instances solved by preprocessing omitted)}
    \label{tab:evaluation:unsat:solvedUnsolved}
\end{wraptable}

Do note, that the cost and contribution must always be considered in relation to the abstraction scheme at hand.
For example, an instance might be solved in approximation step 3, but only because the constraints from approximation steps 1 and 2 are already in place. In this case, the instance might still not be solved in the first step if one were to swap the approximation steps.
We will actually see an example for this kind of interdependence in the subsequent analysis.
This evaluation technique does not allow an independent evaluation of the single approximation steps, but helps to analyse the stengths and weaknesses of an abstraction scheme by decomposing the overall benefits and downsides onto the approximation steps in a sensible way. Furthermore, this technique might show pathways for further improvements.

\paragraph{Analysing the approximation steps.}
\begin{wraptable}[17]{R}{0.5\textwidth}
    \centering
    \raisebox{0pt}[\dimexpr\height-2\baselineskip\relax]{
    \begin{tabular}{l|c|c}
        Benchmark family\footnotemark & \#more & \#less \\ \hline \hline
        \texttt{rw-Noetzli} & \textbf{6} & 0 \\ \hline
        \texttt{bmc-bv-svcomp14} & \textbf{2} & 0 \\ \hline
        \texttt{brummayerbiere2} & \textbf{2} & 1 \\ \hline
        \texttt{calypto} & \textbf{4} & 0 \\ \hline
        \texttt{Sage2} & \textbf{25} & 15 \\ \hline\hline
        \texttt{UltimateAutomizer} & 0 & \textbf{1} \\ \hline
        \texttt{BuchwaldFried} & 0 & \textbf{4} \\ \hline
        \texttt{float} & 0 & \textbf{4} \\ \hline
        \texttt{log-slicing} & 0 & \textbf{3} \\ \hline \hline
        \textbf{Sum} & 39 & 28 
    \end{tabular}
    }
    \caption{Instances solved more/less by Ablector for specific benchmark families in comparison to Boolector}
    \label{tab:evaluation:unsat:families}
\end{wraptable}
Using the data from the experiments we calculated the costs and contributions of each approximation steps in the scheme (Table \ref{tab:evaluation:contributions:fullscheme}).
The quantitative analysis shows that steps 1 and 3 are the most effective approximation steps of the scheme while step 2 seems to cost more then it contributes.
For step 4, however, we noticed that a qualitative analysis can be just as interesting:
While step 4 can no longer solve the benchmark instance \texttt{log-slicing/bvsdiv\_18.smt2}, ensuring $\bvsdiv$ is correctly implemented for bitwidth 18\footnote{Ablector solved this for bitwidth 15 to 17, Boolector for bitwidth 15 to 18}, the abstraction scheme is able to solve \texttt{calypto/problem\_16.smt2} (a sequential equivalence checking problem).
Depending on the use case one might therefore argue that this approximation should stay in place even though it does not change the total number of solved cases.
In a parallelized portfolio approach it might even be interesting for an abstraction scheme to achieve high contributions at a high cost as other solvers can take care of the problems unsolved by the abstraction scheme in parallel.

\paragraph{Modifying the abstraction scheme.}
Based on the results obtained above, we ran an experiment with Ablector omitting approximation step 2 thus only using steps 1,3 and 4. The results of this experiment in comparison to the full abstraction scheme show that there are 20 benchmark instances which only the full abstraction scheme can solve, while 2 instances can only be solved by the version omitting step 2.
\begin{wrapfigure}[15]{R}{0.6\textwidth}
    \centering
    \raisebox{0pt}[\dimexpr\height-2\baselineskip\relax]{
        \includegraphics[width=0.6\textwidth]{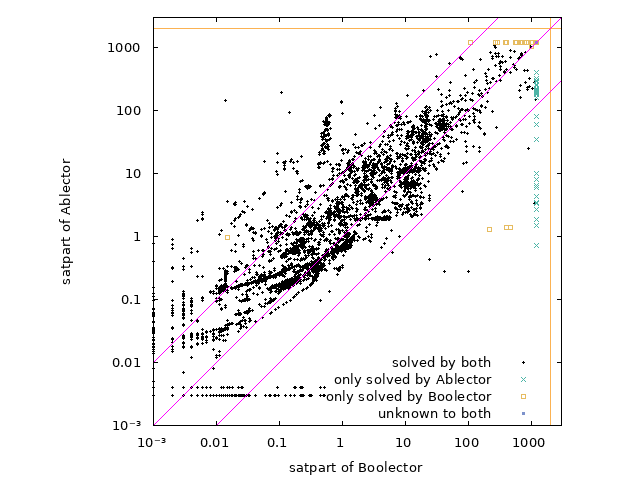}
    }
    \caption{Scatter plot of solving times in Boolector and Ablector (time in seconds)}
    \label{fig:evaluation:unsat:scatter}
\end{wrapfigure}
This seemed to indicate that the subsequent approximation steps rely on the constraints implemented in step 2 in order to function properly, while the constraints alone do not improve the solving behaviour.
This is a classic example of interdependence within the scheme as outlined above.
The full abstraction scheme solves 16 of the 20 mentioned instances in refinement round 3 or later where approximation steps 2, 3 and sometimes 4 are already in use.
This observation motivated a final experiment in which Ablector ran with the previous approximation steps 2 and 3 merged into a single step.
While achieving comparable results in solving times compared to Ablector, this version was able to solve 2 instance the full scheme couldn't solve and only failed at solving one instance the original could have solved.
This evidence supports our hypothesis that the constraints from phase 2 are valuable when combined with the constraints of step 3 and 4 as we do not see the same drop in performace observed for the version which omitted step 2 entirely.

\footnotetext{Names shortened for conciseness}

\paragraph{}
The methodology for abstraction scheme analysis detailed above is a useful tool for analysing schemes for evaluation and optimization purposes.
It further helps in gaining a deeper understanding of the investigated approximations (as seen for step 2).
It will be interesting to see how this methodology can be used to evaluate, understand and optimize other abstraction schemes in the future.

\subsection{On signed and unsigned abstractions}

The approximation steps in our scheme were originally developed when attempting to improve the solving behaviour for instances of the LLBMC family of benchmarks \cite{Iser-Kutzner-Sinz-LLBMC-Bench}.
Considering the \texttt{modmul} benchmark instance in said family, we developed an abstraction scheme differing for signed and unsigned operators.
This is noteworthy, as signed operators are usually just rewritten as unsigned operators during preprocessing.
We experimentally compared our signed abstraction scheme with a version which only considered unsigned operators and rewrote $\bvsdiv$ and $\bvsrem$ to their unsigned versions using if-then-else statements. 
Surprisingly, the results showed no significant difference in runtime or solving behavior on the SMT-LIB benchmark set under evaluation.
This is interesting for two reasons:
On the one hand, the \texttt{modmul} benchmark instance showed that special treatment for signed operators can help for certain benchmark instances while our experiment on the SMT-LIB benchmarks shows that this special treatment does not worsen the overall performance.
On the other hand, the result suggests that the LLBMC family of benchmarks would be an interesting addition to the SMT-LIB benchmark set as it contains instances with novel properties - namely being solvable quite performant using special abstraction schemes for signed operators.

\subsection{Unsatisfiable Benchmarks}
\begin{wraptable}[9]{R}{0.5\textwidth}
    \raisebox{0pt}[\dimexpr\height-1.5\baselineskip\relax]{
    \begin{tabular}{cc|c|c|c}
        &&\multicolumn{2}{c|}{Boolector}&\\
        &&unsolved&solved&\\ \hline
        \multirow{2}{*}{Ablector}&unsolved& 711 & 474 & 1185 \\ \cline{2-5}
        & solved & 47 & 4792 & 4839 \\ \hline
        & & 758 & 5266 & 6024 \\
    \end{tabular}
    }
    \caption{Number of satisfiable instances solved by Boolector and Ablector.}
    \label{tab:evaluation:sat:solvedUnsolved}
\end{wraptable}
We will now look at the overall performance of our abstraction scheme.
Figure \ref{fig:evaluation:unsat:scatter} gives a first overview on the solving times of Ablector in comparison to Boolector.
While Ablector slows down the solving process for a number of instances with short runtime, we see quite a few instances in green at the right side of the scatter plot which can only be solved by Ablector - sometimes even within seconds.
Table \ref{tab:evaluation:unsat:solvedUnsolved} presents a more concise summary of the instances only solved by Ablector or Boolector (omitting instances solved through preprocessing).
Ablector shows a slightly better performance for the total number of solved instances (11 instances more) and solves 39 instances Boolector fails upon.
A large share of the instances solved by both procedures can be considered as easy: Only 182 of the 5904 instances solved by both took longer than 100s for one of the solvers. In comparison to this, an addition of 39 uniquely solved instances is a considerable progress - especially for portfolio approaches and cases where many unsatisfiable instances which are believed to be hard need to be solved.
Table \ref{tab:evaluation:unsat:families} presents evidence that our abstraction scheme contributes to solving some instances of sequential equivalence checking (in \texttt{calypto} \cite{reisenberger2014pboolector}) as well as a number of instances in the \texttt{Sage2} benchmark family \cite{godefroid2008automatedSAGE} concerned with constraint resolution for whitebox fuzz testing. Furthermore, our approach helps with the verification of rewrite rules in the context of \cite{notzli2019syntax} (\texttt{rw\hyp{}Noetzli}).
Apart from above mentioned instances with published unsatisfiability status, Ablector was even able to solve seven instances of the rewrite rule verification family with status \textit{unknown} within the SMT-LIB benchmark set, which Boolector failed to solve\footnote{Boolector was able to solve exactly 40 of the 92 unknown instances through its rewriting engine - Ablector solved another 7 through its abstraction scheme}.

\subsection{Satisfiable Instances}
For satisfiable instances on the other hand, Ablector's performance is visibly worse than Boolector's: As we can see in Table \ref{tab:evaluation:sat:solvedUnsolved}, Boolector is able to solve a lot of instances Ablector cannot currently solve and the runtime Ablector takes for the solved instances cannot make up for this flaw; neither can the 47 instances only solved by Ablector.
While about 500 timed out instances get stuck in the first refinement round, the rest of the timed out instances are evenly distributed across all refinement rounds.
Bounding the running time of each refinement round by an upper limit could potentially avoid the problem of instances getting stuck in a certain step.
At the same time, such a time out must be fine-tuned in a manner which does not break the positive effects of our abstraction scheme for unsatisfiable instances.
We expect that the abstraction scheme's performance could be improved in future work by integrating the abstractions directly into a solver like Boolector instead of building them as a layer on top. This would allow to make better use of already implemented under-approximation techniques that are completely ignored for most abstraction steps in the current scheme.

\section{Conclusion}
\label{ch:Conclusion}
We introduced an approach for solving quantifier-free bit-vector problems in SMT-LIB's \texttt{QF\_BV} theory. The approach is based on abstraction methodologies previously used for various other problems in logic and specifically in SMT.
We presented numerous approximation steps for 5 comparatively costly functions of the bit-vector theory. Additionally, we gave both a theoretical definition of such abstraction schemes and presented a methodology allowing the experimental analysis of single approximation steps within a given abstraction scheme.

We saw that the presented approach performs better than Boolector in deciding unsatisfiable bit-vector problems, solving 11 unsatisfiable instances and 7 instance with unknown status more and yielding a total of 46 uniquely solved instances with unsatisfiable or unknown status in comparison to Boolector. However, the implemented prototype is not yet competitive for satisfiable instances. This is in some way a natural result, as over-approximations usually improve the solver runtime on unsatisfiable (and not on satisfiable) instances \cite{Brummayer-PhD}. Also, the difference in solved instances for both unsatisfiable and satisfiable problems makes Ablector a promising addition for a portfolio solver.

As already seen in UCLID \cite{Bryant2007_Chapter_DecidingBit-VectorArithmeticWi-UCLID} interleaving over- and under-approximations has the potential to yield a solver which solves satisfiable and unsatisfiable benchmark instances equally well - this could also be an option for the abstraction scheme presented here.
However, well-tuned time limits or intelligent interruption conditions, possibly based on Luby Sequences \cite{luby1993optimal}, will be necessary for all approximation steps in order to avoid \textit{lock ins} where the solver keeps working in a single phase without coming to any result, while still granting the steps enough time to come to conclusions where possible.
Alternatively, a portfolio approach making use of various over- and under\hyp{}approximation tech\-niques could be explored.
 
\bibliographystyle{splncs03}
\bibliography{main}

\appendix
\pagebreak
\section{Correctness of the Abstraction Approach}
\label{sec:appendix:proof}
This section complements Section \ref{ch:Scheme}.
First, we provide a proof for the completeness of $\mathcal{C}_{\texttt{op}}\left(\overline{x}\right)$. Afterwards, we explain how a model for some $\phi$ can be constructed given a model for $\phi$'s abstraction and vice-versa.

In the following, we assume that a $T+UF$-interpretation for some theory $T$ is also a $T$ interpretation where the evaluation for the uninterpreted functions is ignored. Furthermore, we assume that we can extend a $T$-interpretation into a $T+UF$-interpretation by adding evaluations for the necessary uninterpreted functions. This can usually be considered as valid  (e.g. a $\qfufbv$ model of some formula $\phi$ can also be a $\qfbv$ model of $\phi$ if $\phi$ does not contain any undefined functions).

\begin{lemma}[Completeness of Abstraction Schemes]
   \label{lemma:Scheme:Abstraction_Completeness}
Given some $T$-abstrac\-tion scheme 
$\mathcal{AS}_{\texttt{op}} = \{ \left(ap_{\texttt{op}}, \mathcal{A}^1_{\texttt{op}}\right),\dots,\left(ap_{\texttt{op}}, \mathcal{A}^k_{\texttt{op}}\right) \}$
with the properties defined above,
$\mathcal{C}_{\texttt{op}}$ is a complete $T$-approximation of \texttt{op}.
\begin{proof}
Let $\overline{x}$ be an arbitrary input vector for \texttt{op}.
For any $T+UF$-interpretation $\mathcal{I}$ with $\mathcal{I} \vDash ap_{\texttt{op}}\left(\overline{x}\right) \doteq \texttt{op}\left(\overline{x}\right)$, we know that by definition $\mathcal{I} \vDash \mathcal{A}^i_{\texttt{op}}\left(\overline{x}\right)$ for all $i\in\llbracket1,k\rrbracket$ as all approximations $\mathcal{A}^i_{\texttt{op}}$ are complete. 
Therefore,
\[
   \mathcal{I} \vDash \bigcup\limits_{\left(\wc,\mathcal{A}\right)\in\mathcal{AS}_{\texttt{op}}} \mathcal{A}\left(\overline{x}\right)
   \text{ (i.e., }\mathcal{I} \vDash \mathcal{C}_{\texttt{op}}\left(\overline{x}\right)\text{)}\enspace,
\]
which implies that $\mathcal{C}_{\texttt{op}}$ is a complete $T$-approximation, too. 
\end{proof}
\end{lemma}

\begin{theorem}[Correctness of Abstraction Approach]
   \label{theorem:Scheme:Equivalence}
   Let $T=\left(\Sigma,A\right)$ be some theory with $\texttt{op}\in\Sigma^F$, $\mathcal{s}^F\left(\texttt{op}\right)=\sigma_1\dotsi\sigma_n\sigma$ and $n\geq1$. 
   Let further $\Phi$ be an arbitrary $\Sigma$-formula containing some function application $\texttt{op}\left(\overline{x}\right)$.
   For any $T$-abstraction scheme $\mathcal{AS}_{\texttt{op}}$ with function symbol $ap_{\texttt{op}}$, the following property holds:\\
   There exists a $T$-interpretation $\mathcal{I}_{\Phi}$ which is a $T$-model for $\Phi$ iff there exists a $T+UF$-interpretation $\mathcal{I}_{\mathcal{A}}$ which is a $T+UF$-model for 
   \[
   \Psi \coloneqq \Phi\left[ \texttt{op}\left(\overline{x}\right) \mapsto ap_{\texttt{op}}\left(\overline{x}\right) \right] \land \bigwedge\limits_{\left(\wc,\mathcal{A}\right)\in\mathcal{AS}_{\texttt{op}}} \mathcal{A}\left(\overline{x}\right) \,.
   \]
   \begin{proof}
   The theorem will be proven in two directions. For each direction, we will construct a suitable interpretation given the premise interpretation.
   \begin{itemize}
       \item[$\Rightarrow$] Let $\mathcal{I}_{\Phi}$ be a $T$-model for $\Phi$.
           We build a model $\mathcal{I}_{\mathcal{A}}$ by extending $\mathcal{I}_{\Phi}$ so that $\mathcal{I}_\mathcal{A}\left(ap_{\texttt{op}}\left(\overline{x}\right)\right)$
           evaluates to $\mathcal{I}_{\Phi}\left(\texttt{op}\left(\overline{x}\right)\right)$. This is possible as $ap_{\texttt{op}}$ is a new uninterpreted function symbol not used within $\Phi$.
           As $\mathcal{I}_{\mathcal{A}} \vDash ap_{\texttt{op}}\left(\overline{x}\right) \doteq \texttt{op}\left(\overline{x}\right)$,
           the completeness proof in Lemma \ref{lemma:Scheme:Abstraction_Completeness} yields
           $\mathcal{I}_{\mathcal{A}} \vDash \mathcal{A}_{\texttt{op}}\left(\overline{x}\right)$.
           Therefore 
           $\mathcal{I}_{\mathcal{A}} \vDash\Phi\left[ \texttt{op}\left(\overline{x}\right) \mapsto ap_{\texttt{op}}\left(\overline{x}\right) \right] \land \bigwedge_{\left(\wc,\mathcal{A}\right)\in\mathcal{AS}_{\texttt{op}}} \mathcal{A}\left(\overline{x}\right)$. \\            
       \item[$\Leftarrow$]  Let $\mathcal{I}_{\mathcal{A}}$ be a $T+UF$-model for $\Psi$.
           The abstraction scheme definition states that
           $\mathcal{I}_{\mathcal{A}} \vDash \bigwedge_{\left(\wc,\mathcal{A}\right)\in\mathcal{AS}_{\texttt{op}}} \mathcal{A}\left(\overline{x}\right)$
           implies $\mathcal{I}_{\mathcal{A}} \vDash ap_{\texttt{op}}\left(\overline{x}\right) \doteq \texttt{op}\left(\overline{x}\right)$ through the soundness property.
           This implies that $\mathcal{I}_{\mathcal{A}} \vDash \Phi$.
   \end{itemize}
   \end{proof}
\end{theorem}

Finally, the following lemma shows that the sole requirement for a correct abstraction scheme is that all over-approximations must be implications of the original function while the set of all approximations in the scheme must be an implicant of the original function:
\begin{lemma}[Soundness/Completeness through Implication]
   \label{lemma:Scheme:Implication}
   Given some theory $T=\left(\Sigma,A\right)$ and a $T$-approximation $\left(ap_{\texttt{op}}, \mathcal{A}_{\texttt{op}}\right)$. 
   If for all $T+UF$-interpretations $\mathcal{I}$ and all $\overline{x}\in Dom\left(\mathcal{A}_{\texttt{op}}\right)$
   \[
       \mathcal{A}_{\texttt{op}}\left(\overline{x}\right) \implies ap_{\texttt{op}}\left(\overline{x}\right) \doteq \texttt{op}\left(\overline{x}\right)
   \]
   holds, then $\left(ap_{\texttt{op}}, \mathcal{A}_{\texttt{op}}\right)$ is sound.

   If for all $T+UF$-interpretations $\mathcal{I}$ and all $\overline{x}\in Dom\left(\mathcal{A}_{\texttt{op}}\right)$
   \[
       ap_{\texttt{op}}\left(\overline{x}\right) \doteq \texttt{op}\left(\overline{x}\right) \implies \mathcal{A}_{\texttt{op}}\left(\overline{x}\right)
   \]
   holds, then $\left(ap_{\texttt{op}}, \mathcal{A}_{\texttt{op}}\right)$ is complete.
   \begin{proof}
       The proof is based on Definitions \ref{def:Scheme:Sound} and \ref{def:Scheme:Complete}.\\
       Given for some formula, the soundness (completeness) formula above holds for all $\mathcal{I}$ and $\overline{x}$:\\
       For any interpretation $\mathcal{I}$ where
           $\mathcal{I}\nvDash\mathcal{A}_{\texttt{op}}\left(\overline{x}\right)$
           ($\mathcal{I}\nvDash ap_{\texttt{op}}\left(\overline{x}\right) \doteq \texttt{op}\left(\overline{x}\right)$)
           the definition for soundness (completeness) is already fulfilled.\\
       In case
           $\mathcal{I}\vDash\mathcal{A}_{\texttt{op}}\left(\overline{x}\right)$
           ($\mathcal{I}\vDash ap_{\texttt{op}}\left(\overline{x}\right) \doteq \texttt{op}\left(\overline{x}\right)$) for some interpretation $\mathcal{I}$  ,
       then we know through the formula above that
           $\mathcal{I}\vDash ap_{\texttt{op}}\left(\overline{x}\right) \doteq \texttt{op}\left(\overline{x}\right)$
           ($\mathcal{I}\vDash\mathcal{A}_{\texttt{op}}\left(\overline{x}\right)$)
       which implies that the approximation is, by definition, sound (complete).

   \end{proof}
\end{lemma}
\section{Reproducibility}
\label{sec:appendix:reproducibility}
\subsection{Software}
For alle experiments a modified version of \texttt{Boolector 3.2.0} is used.
More specifically, we modified commit \texttt{ec1e1a93\-21aac2\-5e22\-d404\-368fef0\-52f70\-4ce78b} so that we could measure the time of the \texttt{check-sat} instruction. This can be found under
\url{https://github.com/samysweb/boolector} 
in branch \texttt{sat-time\-measure-32}.
As underlying SAT-solver Lingeling with version
\texttt{bcj 78ebb8672540b\-de\-0a335\-aea946bbf32515157d5a} is used.
All software packages were compiled using the provided cmake scripts which have the highest optimization levels enabled using gcc in version \texttt{(Ubuntu 7.5.0-3ubuntu1~18.04) 7.5.0}.
For the final experiments presented, Ablector is used in the version available in commit \texttt{45fc7e1\-c388ba92f\-7f34\-e8f\-571ad109\-a1c7eb240} at\\
\url{https://github.com/samysweb/ablector}.
In our experiments we used a version of Ablector which used a new function symbol for each function application during the first 2 phases of abstraction.
This was done as this version showed slightly more promising results than the version which reused the same function symbol.

\subsection{Machine}
All experiments were executed on a cluster of 20 identical compute nodes each housing 2 Intel Xeon E5430 @ 2.66GHz CPUs and a total of 32GB of RAM.
The SMT benchmark files were stored on a RAID system connected to the cluster.

\subsection{Benchmark execution}
Two jobs were run in parallel on each compute node with the timeout set to 1200 seconds
this posed no caching issues as they were run on seperate CPU sockets
\footnote{Early on we ran up to 8 experiments on a single node to make use of the available cores however this seemed to produce caching issues slowing down the experiment times.}.
For time surveillance and measurements we used the runlim utility.
All benchmarking scripts and the log results can be obtained at 
\url{https://github.com/samysweb/BA-experiments}.

\end{document}